\documentstyle[aps,twocolumn,epsfig,eqsecnum]{revtex}

\begin{document}

\newcommand{\mubf}{\mbox{\boldmath $\mu$}}

\twocolumn[\hsize\textwidth\columnwidth\hsize\csname    
@twocolumnfalse\endcsname                               

\begin{title} {\Large  \bf
Hexatic$-$Herringbone Coupling at the Hexatic Transition in Smectic 
Liquid Crystals: $4-\epsilon$ Renormalization Group Calculations 
Revisited}
\end{title} 

\author{Mohammad Kohandel$^1$, Michel J.P. Gingras$^{1,2}$ and Josh P. Kemp$^1$}
\address{$^1$Department of Physics, University of Waterloo, 
Ontario N2L 3G1, Canada} 
\address{$^2$Canadian Institute for Advanced Research,
180 Dundas Street West,  Toronto, Ontario,  M5G 1Z8, Canada}

\date{\today} 
\maketitle 

\begin{abstract}
{\noindent Simple symmetry considerations would suggest that 
the transition from the smectic-A phase to the long-range bond
orientationally ordered hexatic 
smectic-B phase should belong to the XY universality 
class.
However, a number of experimental studies have constantly 
reported over the past twenty years ``novel'' critical behavior
with non-XY critical exponents for this transition. Bruinsma and Aeppli argued in 
Physical Review Letters {\bf 48}, 1625 (1982), using a 
$4-\epsilon$ renormalization-group calculation, that 
short-range molecular herringbone correlations coupled to the 
hexatic ordering drive this transition first order via thermal 
fluctuations, and that the critical behavior observed 
in real systems is controlled by a `nearby'
tricritical point.  We have 
revisited the model of Bruinsma and Aeppli and present here the 
results of our study. We have found two nontrivial 
strongly-coupled herringbone-hexatic fixed points apparently 
missed by those authors. Yet, those two new 
nontrivial fixed-points
are unstable, and we obtain the same final conclusion as the 
one reached by Bruinsma and Aeppli, namely that of a 
fluctuation-driven first order transition. We also discuss
the effect of local two-fold distortion of the bond order as a
possible missing order 
parameter in the Hamiltonian.}

\end{abstract}

\vskip2pc]                                              


\newpage


\section{Introduction}

The nature of phase transitions
in two dimensional (2D) systems, has been the subject
of numerous investigations over the last three decades. 
According to the
Mermin-Wagner-Hohenberg theorem \cite{MWH}, 
the continuous symmetry
of the XY and Heisenberg models cannot be spontaneously
broken at finite temperature, and there can be no long
range magnetic order. However, Kosterlitz and Thouless
(KT) \cite{KT73} argued that there is a new type of phase
transition from a high temperature phase with exponentially
decay of the correlations to a low temperature
phase with power law decay of the correlations.
The idea of KT has been extended by Halperin and Nelson 
\cite{HN78} and Young \cite{Young79} to the 2D melting
problem. One of the main results of the KTHNY theory
is the prediction of an intermediate 2D phase called
the hexatic phase for systems that have
a six-fold (hexagonal) symmetry in their
crystalline ground state. 
This hexatic phase displays short range positional order, 
but quasi long range bond-orientational 
order, which
is different from the true long range bond-orientational
and quasi long range 
translational order  of a 2D solid phase \cite{HN78,Strandburg}. 
The hexatic phase can be characterized
by a bond-orientational order parameter defined
 by $\Psi_6=\mid\!\Psi_6\!\mid\exp(i6\psi_6)$.
Assuming that the hexatic state occurs and is not preempted
by a direct first order transition from the solid to the
isotropic liquid phase, the 
system should, in the simplest scenario for 2D, display either a KT transition
or a first order transition from the hexatic state to the
isotropic liquid phase
\cite{Strandburg}.

It was soon realized after the proposal of the 
KTHNY theory that the hexatic phases, with
short range positional order but true long range
bond orientational order, might exist in highly
anisotropic three dimensional (3D) systems. 
Specifically, Birgeneau and Lister \cite{BL78} applied the 
notion of a hexatic state of the 2D 
melting theory to 3D liquid crystal phases consisting 
of stacked 2D layers. They proposed
that some of the experimentally observed smectic liquid
crystal phases could be physical realization of
3D hexatics. Birgeneau and Lister suggested that the (weak) 
interlayer interaction could promote
the quasi long range 
order of 2D hexatic layers 
to true long range bond orientational order in 3D.

Stimulated by 
these theoretical advances, numerous experimental 
efforts have been undertaken to test theoretical
predictions in
different liquid crystal materials candidate for
displaying hexatic phases \cite{HuangRvs}. 
An x-ray study of the liquid crystal compound 65OBC
(n-alkyl-4'-m-alkoxybiphenyl-4-carboxylate,n=6,m=5)
\cite{Pindak81} provided the first
indication of the existence of the 3D analog of
the 2D hexatic phase. It was also 
found that in addition to the hexagonal
pattern of diffuse spots of scattered intensity, which is the signature
of the hexatic phase, there are some broader 
peaks corresponding to correlations in the
molecular orientations about their long axes 
\cite{Pindak81}. The positions of these peaks show 
that, locally, the molecules are packed according 
to a herringbone pattern perpendicular to the 
smectic layer stacking direction (see Fig. 1). Despite the indication
of short range herringbone correlations, this novel phase 
is simply 
denoted as the hexatic-B (HexB) phase. Upon 
increasing temperature, this phase looses its
long range bond orientational order and undergoes a
transition to the 
smectic-A (SmA) phase, which 
essentially consists of a stack of 2D liquid layers.
Upon cooling, the HexB phase transforms 
via a first order phase transition into the 
crystal-E (CryE) phase, which exhibits both
long range translational order and long range herringbone 
orientational order in the orientations of the molecular axes.
\begin{figure}
\begin{center}
\rotatebox{0}{\includegraphics[width=6cm]{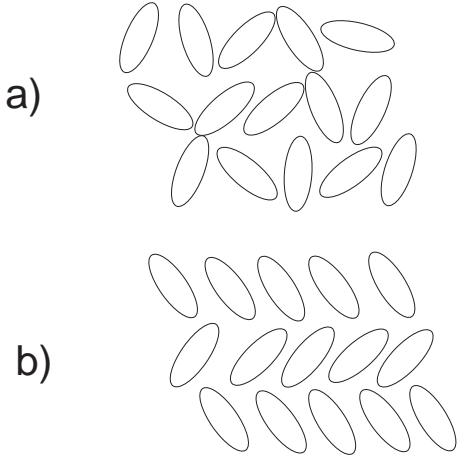}}
\vspace{3mm}
\caption{Local hexagonal coordination of the molecules `as seen'
along the stacking direction of the smectic layers.
The elliptical shape of the molecules as seen along the 
stacking direction is meant to represent the ``wide''
benzene rings present on most thermotropic liquid crystal
molecules. 
a) without herringbone correlations. 
b) with local
herringbone
packing correlations of the molecular axes.}
\end{center}
\end{figure}
According to the $U(1)$ symmetry 
of the $\Psi_6$ bond orientational order parameter, one
would naively 
expect to find XY-like critical exponents at
the
SmA$-$HexB transition.
However, heat capacity investigations
near the SmA$-$HexB transition of 65OBC 
\cite{HuangRvs,Pindak81,Huang81} and subsequent 
calorimetric studies on many other components 
in the nmOBC homologous series \cite{HuangRvs,Pitchford85} 
have constantly been reporting continuous 
(second order) SmA$-$HexB transitions with 
very large values for the heat capacity critical 
exponent, $\alpha\approx 0.6$. This is drastically
different from the 3D 
XY critical exponent 
$\alpha=-0.007$ \cite{LZ85}.
As well, thermal conductivity and birefringence experiments
have allowed the determination of other static critical exponents,
all of which differ systematically from the 3D XY value, while
they, together, obey the standard scaling relationships
expected for a genuine second order phase transition \cite{HuangRvs}.

In the light of the existence of the short range 
herringbone fluctuations, detected in the x-ray 
diffraction studies \cite{Pindak81}, Bruinsma
and Aeppli (BA) \cite{BA82} formulated a 
Ginzburg-Landau theory that
includes both the hexatic and herringbone order.
Because the HexB phase exhibits only short range 
positional order, BA suggested that the herringbone
order can also be represented by an XY order
parameter described by 
$\Phi_2=\mid\!\Phi_2\!\mid\exp(i2\phi_2)$. 
At the microscopic level, it is the
molecular anisotropy that creates
a coupling between the hexatic bond order and the
herringbone molecular order~\cite{Gingras,Gingras2}.
At the phenomenological Ginzburg-Landau level, this
coupling is described by a hexatic-herringbone
interaction term
$V_{\rm hex-her} = h {\rm Re}(\Psi_6^*\Phi_2^3)$.

BA constructed an appropriate free energy density 
based on symmetry considerations and investigated 
the effects of
fluctuation corrections to the mean
field behavior for 3D systems. In the mean field
approach, their results indicate that the SmA$-$HexB 
transition should be continuous. 
However, $4-\epsilon$ renormalization group (RG) 
calculations, which includes thermal fluctuations 
and the coupling term $h {\rm Re}(\Psi_6^*\Phi_2^3)$, 
show that short-range molecular herringbone 
correlations coupled to the hexatic ordering 
drive this transition first order, which becomes 
second order at a tricritical point\cite{BA82}. 

Interestingly,  heat capacity measurement studies
of (truly two-dimensional)
two-layer free standing films of different nmOBC 
compounds yield very sharp heat capacity peaks
near the SmA$-$HexB transition which can be described 
by the critical exponent $\alpha\approx 0.3$
\cite{HuangRvs,Stoebe}.This is in sharp contrast with
the usual broad and nonsingular specific heat hump associated
to the KT transition in the 2D XY model, or yet the
first order transition that could occur in a physical
system where the vortex core energy is less than some
critical value\cite{Strandburg,Minnhagen}. This $\alpha \approx 0.3$ result
further suggests that the SmA$-$HexB cannot
be described by a simple model with a unique
(critical) XY-like order parameter.
In this context,
there has been some numerical simulations done
to obtain more insight into the nature of the 
SmA$-$HexB transition in 2D systems. 
The model used in the simulations 
\cite{Huang93,Huang96} consists of a 2D 
  lattice of coupled XY spins 
based on the BA Hamiltonian. 
The simulation results suggest the existence
of a new type phase transition in which the two 
different orderings are simultaneously established
through a continuous transition. It is interesting to
note here that, in seemingly different context,
there has also been numerous theoretical and 
numerical attempts to identify `novel chiral'
universality classes for 
systems such
as frustrated XY model and Ising-XY coupled model
\cite{Kosterlitz}.

Certainly, for three dimensions,
the scenario of a fluctuation-driven first order SmA$-$HexB transition
due to hexatic$-$herringbone coupling would appear reasonable for the
SmA$-$HexB transition in 65OBC which, upon further cooling, undergoes a
HexB$-$CryE transition that establishes long range herringbone and
positional order. However, the mixture of 3(10)OBC and PHOAB exhibits a
very large temperature range for the HexB phase above the
crystallization temperature to the CryE phase. If there were
herringbone fluctuations near the SmA$-$HexB transition in that
mixture, one could expect them to be quite small because of the large
HexB temperature range, and the  SmA$-$HexB transition could then
possibly be continuous, and to belong to the (then naively expected) XY
universality class.  However, the fact that the SmA-HexB
transition in the 3(10)OBC-PHOAB mixture is found to be first order
does not support this simple minded argument \cite{HuangRvs}.
Following the same type of reasoning,
 recent x-ray diffraction studies on 75OBC \cite{HuangRvs} show that
the intensity of the herringbone peaks is weaker than those of 65OBC.
In principle, if one assumes that 65OBC is near a tricritical point,
75OBC should therefore be further removed from this point (due to the
weaker herringbone diffraction peaks, and consequently, weaker
$V_{\rm hex-her}$), with again the possibility to
recover 3D XY critical behavior.  Yet, the same (unconventional)
heat-capacity critical exponents are found for these two materials.

The experimental results above could 
be interpreted as
a possible indication
of an underlying ``novel'' (non-XY) stable
fixed point that control the SmA$-$HexB transition when herringbone
correlations are present.
The apparent lack of progress on the theoretical side
of the SmA$-$HexB problem has led us 
to reinvestigate the model of 
BA and to, specifically, look for a possible calculation
error. Firstly, it is important to note that the
conclusion of a fluctuation-driven first order within
a $4-\epsilon$ calculation is acutely depending on
the `numerics' and not constrained by symmetry consideration:
a small error (such as a factor 2 instead of 4 here or there)
can change the conclusion of a fluctuation-driven first
order transition. 
Secondly, and more specific to the BA problem, 
we show in the next section, when describing
the Ginzburg-Landau free energy 
density for the SmA$-$HexB transition, 
that 
some terms in the RG equations, 
to first order in  $\epsilon$, were missed in 
the work of BA. Based on our RG equations,
we find two nontrivial strongly-coupled 
herringbone-hexatic fixed points, apparently 
missed by those authors. However, 
those two nontrivial fixed-points are unstable, 
and we reach the same final conclusion as the 
one found by BA, namely that of a 
fluctuation-driven first order transition. 
We also discuss the possibility of a third and
a priori
possibly physically pertinent
order parameter in the Hamiltonian model
of the  SmA$-$HexB transition.
Because of local distortion of the bond
orientational order induced by
the anisotropy of the intermolecular 
potential 
\cite{Gingras,Gingras2}
and the  
herringbone correlations, one may
generalize the Hamiltonian to the 
case with three XY-like order parameters,
in which two of them are two-fold symmetric,
one for the herringbone correlations, $\Phi_2$,
and one for the
local two-fold distortion, $\Psi_2$,
and a third order parameter
with six-fold symmetry, $\Psi_6$, related
to hexatic ordering. We discuss both
mean field and RG calculations for this
new model.

The rest of this paper is organized as follows.
In Sec.~II.A, we reintroduce the BA model and
present the result of our RG calculations. In the
Sec.~II.B, we generalize the Hamiltonian to
the case of three order parameter and discuss
the mean field theory and RG results.  
The discussions and conclusions appear in Sec.~III. 

\section{Models and RG Calculations}

\subsection{BA Hamiltonian}

To formulate the Landau Ginzburg (LG) free energy, 
which describes both the hexatic and herringbone order,
one recalls that the hexatic order is six-fold
symmetric, while rotating a herringbone pattern
by $180^o$ leaves it unchanged. Consequently, the
appropriate 
LG free energy ought to be invariant with respect
to the transformation $\phi_2(r)\rightarrow\phi_2(r)+n\pi$
and $\psi_6(r)\rightarrow\psi_6(r)+m(2\pi/6)$ where
$n$ and $m$ are integers. Thus to lowest order
in $\Psi_6$ and $\Phi_2$, the BA Hamiltonian~\cite{BA82}
is:
\begin{eqnarray}\label{LGF1}
\beta F&=&\int d^3x \; 
\bigg[
\frac{r_6}{2}\mid\!\Psi_6\!\mid^2
+
\frac{1}{2}  \mid\!\nabla\Psi_6\!\mid^2
+\frac{r_2}{2}\mid\!\Phi_2\!\mid^2 \nonumber  \\
&+&\frac{1}{2}\mid\!\nabla\Phi_2\!\mid^2
+u_6\mid\!\Psi_6\!\mid^4+u_2\mid\!\Phi_2\!\mid^4 \nonumber \\
&+&w\mid\!\Phi_2\!\mid^2\mid\!\Psi_6\!\mid^2
+h {\rm Re}(\Psi_6^*\Phi_2^3)\bigg].
\end{eqnarray}
The condition for thermodynamic stability of $F$ 
for $w=0$ is $h^{4/3}<(4^{4/3}/3)u_2 u_6^{1/3}$.
This condition can be obtained by minimizing the free 
energy density on the critical isothermal
line $r_2=r_6=0$ and requesting that $\beta F>0$. 
As discussed in the Ref. \cite{BA82}, in the mean field
approximation, for $w=0$ and $h=0$ the phase
diagram in the $r_2-r_6$ plane includes four
distinct phases: an isotropic (SmA) phase with
$\Psi_6=\Phi_2=0$, a hexatic (HexB) phase with
no herringbone order with
$\Psi_6\neq 0, \Phi_2=0$,
a ``putative''  `herringbone liquid crystal'
phase with $\Psi_6=0, \Phi_2\neq 0$, and a fully ordered
state with $\Phi_2\neq 0, \Psi_6\neq 0$ akin to a crystalline
E phase\cite{cryE_crystal}, with all these phases separated by
second order transitions. However, if $h\neq 0$,
the herringbone liquid crystal state with no hexatic
order ($ \Phi_2\neq 0, \Psi_6=0$) is 
eliminated because $\Phi_2$ acts as a symmetry
breaking field on $\Psi_6$.
Within mean field theory, the transition lines
between the isotropic and ordered phases remain second
order for $h\neq 0$, and terminate together with
the first order line at a multicritical point\cite{locked-hexatic_nematic}.

We now discuss the RG flow equations and the stability 
of the fixed points (FPs). Our calculations show
that the RG equations to first order in   
$\epsilon=4-d$ are:
\begin{eqnarray}\label{RGE}
&&\frac{dr_2}{dl}=2r_2+\frac{16 K_4 u_2}{1+r_2}
+\frac{4K_4w}{1+r_6} \nonumber \\  
&&\frac{dr_6}{dl}=2r_6+\frac{16 K_4 u_6}{1+r_6}
+\frac{4K_4w}{1+r_2}  \nonumber  \\
&&\frac{du_2}{dl}=\epsilon u_2-40K_4 u_2^2-2K_4w^2
-9K_4h^2 \nonumber  \\
&&\frac{du_6}{dl}=\epsilon u_6-40K_4 u_6^2
-2K_4w^2   \nonumber  \\
&&\frac{dw}{dl}=\epsilon w-16K_4 w u_2-16K_4 w u_6
-8K_4w^2-18K_4h^2   \nonumber  \\
&&\frac{dh}{dl}=\epsilon h-24K_4 h u_2-12K_4 h w ,
\end{eqnarray}
where $K_4=1/8\pi^2$. 
The above RG equations differ from those
found by BA in Ref.\cite{BA82}:

\begin{itemize}
\item  The 
first set of  differences 
are the $4  K_4w/(1+r_6)$ and 
$4  K_4w/(1+r_2)$ terms 
in the first and second equations, while
BA have
$2  K_4 w /(1+r_6)$ 
and $2  K_4 w /(1+r_6)$. The extra factor $2$ comes
from the fact 
that the fields $\Psi_6$ and $\Phi_2$ are complex
and the related correlations have two components.
For $h=0$, Eq.~(\ref{RGE}) (with the factor $4 K_4 w$)
reproduces the RG equations of coupled two components
 two-vector model
as in previous studies\cite{Aharony,Brezin}.
We therefore believe that the above RG equations 
for $dr_2/dl$ and $dr_6/dl$ are correct.

\item
Compared to the BA equations, we also obtain two completely
new 
terms, $-18K_4 h^2$,
 in 
the fifth equation and $-12 K_4 h w$ in the sixth equation, 
which can be easily checked using Feynmann diagram 
technique.  Indeed, these two new terms come
from the connected diagrams in the second
order perturbative RG obtained by multiplication
of the relevant diagrams of 
$h \Psi_6^*\Phi_2^3$
with $ h \Psi_6\Phi_2^{*3}$,
and
of $w\mid\!\Phi_2\!\mid^2\mid\!\Psi_6\!\mid^2$
with $h {\rm Re}(\Psi_6^*\Phi_2^3)$, respectively.

\end{itemize}

Because of the two extra terms in the RG equations
for $dw/dl$ and $dh/dl$, we obtain, in
addition to the simple decoupled FP ($r_6^*= r_2^*=
-\epsilon/5, u_6^*=u_2^*=\epsilon/(40K_4)$, $w^*=h^*=0$), 
two new fixed points such that 
$(w^*\neq 0,h^*=0)$ and ($w^*\neq 0,h^*\neq 0)$. 
The first nontrivial FP is given by $h^*=0, r_6^*= 
r_2^*=-\epsilon/4, u_6^*=u_2^*=\epsilon/(48K_4)$, 
and
w$^*=\epsilon/(24K_4)$. This FP, akin to the one
found in minimally coupled two component two vector
 model
\cite{Aharony,Brezin}, was not discussed by BA.

However, and most interestingly, we find 
another ``new'' nontrivial mixed herringbone-hexatic
FP  with all the couplings 
being non-zero:
\begin{eqnarray}
\label{FP2}
r_6^* & = & -0.24845566\epsilon, \nonumber \\
r_2^* & = & -0.24018995\epsilon,	 \nonumber \\   
u_6^* & = & 0.01941403\epsilon/K_4,		 \nonumber \\   
u_2^* & = & 0.01838082\epsilon/K_4,		 \nonumber \\   
w^*   & = & 0.04657169\epsilon/K_4, 		 \nonumber \\   
h^*   & =& \pm 0.00766519\epsilon/K_4.		 \nonumber \\   
\end{eqnarray}
Therefore, based on our RG calculations, there is a FP
with $h^* \neq 0$, which was not found in the previous work of BA.
Linearizing the recursion relations in the vicinity of the FPs 
yields for the FP with ($w^*\neq 0$, $h^*=0$):
$y_1=2-\epsilon/2, y_2=2-\epsilon/6, y_3=-\epsilon,
y_4=-2\epsilon/3, y_5=y_6=0$. Thus, two eigenvalues
are marginal, compatible with what has been
found in similar minimally coupled two component two vector
model \cite{Aharony}.
The eigenvalues for the FP with ($h^*\neq 0,w^*\ne 0)$ above
are 
\begin{eqnarray}
\label{EigenV}
y_1 & = & 2-0.488829\epsilon,	 \nonumber \\
y_2 & = & 2-0.115889\epsilon,	 \nonumber \\
y_3 & = & -0.997894\epsilon,  \nonumber \\
y_4 & = & -0.537266\epsilon, \nonumber \\
y_5 & = & +0.121467\epsilon,  \nonumber \\
y_6 & = & +0.0402392\epsilon.
\end{eqnarray}
These results show that there are four positive  eigenvalues, and
 the above ``new'' nontrivial FP is therefore unstable. 
The two largest (most positive) eigenvalues, $y_1$ and $y_2$, correspond, respectively,
to the thermal eigenvalue and the `relative' coupling strength that 
positions the system in coupling parameter space and determines
what sequence of phase transition occurs. Namely, 
 isotropic$\rightarrow$(hexatic+herringbone) 
via a unique phase transtion or
isotropic $\rightarrow$ hexatic $\rightarrow$(hexatic+herringbone) via
two distinct phase transitions.
The four eigenvalues $y_3$$-$$y_6$ essentially control
the flow in the $w-h$ plane. $y_5$ and $y_6$  are positive, and 
we interpret the result
as an indication that the
above ``new'' nontrivial mixed herringbone-hexatic
FP is unstable. 
Indeed,
we have confirmed this by explicit numerical integration of the
RG equations, and found that
the RG flow goes
to the unstable 
region identified above, which we  interpret as the transition being
driven first order by fluctuations. Therefore, while we have
 indeed found some discrepancies
between our RG equations and those of BA, and 
recovered two extra coupled fixed points, we at the end still
reach the same physical conclusion of BA, namely that 
of a fluctuation-driven
first order smA$-$HexB transition.

\subsection{Generalized Hamiltonian}

We expect physically that the local molecular anisotropy (e.g. from the anisotropic
nature of benzene rings found in most thermotropic
liquid crystal materials) to couple to the local bond direction,
and to create a local two-fold distortion of the otherwise perfect
local six-fold symmetric
nearest-neighbor bond order\cite{Gingras,Gingras2}.
Consequently, 
we now discuss the LG free energy, which describes both the 
hexatic and herringbone order, as well as the local two-fold
distortion
of the bond order. If we assume that the distortion of
lattice has two-fold symmetry, with the order parameter
$\Psi_2=\mid\!\Psi_2\!\mid\exp(i2\psi_2)$, then the 
resulting free energy is invariant under the transformation 
$\phi_2(r)\rightarrow\phi_2(r)+n\pi$, 
$\psi_6(r)\rightarrow\psi_6(r)+m(2\pi/6)$, and
$\psi_2(r)\rightarrow\psi_2(r)+p\pi$ where
$n$, $m$, and $p$ are integers. 

From an RG point of view, 
our motivation to expand the symmetry of our Hamiltonian stems
from 
the observation that in $N$ coupled 2-vector models
a new stable fixed points (called mixed
fixed point~\cite{Aharony,Brezin}) appears in the coupling parameter space
(when $N>2$). 
Thus, to lowest order
in $\Psi_6$, $\Phi_2$, and $\Psi_2$, we have $\beta F=\beta F_0
+U$, where the Gaussian part is given by

\begin{eqnarray}\label{LGF0}
\beta F_0&=&\frac{1}{2}\int d^3x\; \bigg[
r_6\mid\!\Psi_6\!\mid^2+r_2\mid\!\Phi_2\!\mid^2
+\tilde r_2\mid\!\Psi_2\!\mid^2  \nonumber  \\
&+&2 r {\rm Re}(\Phi_2\Psi_2^*)+\mid\!\nabla\Psi_6\!\mid^2
+\mid\!\nabla\Phi_2\!\mid^2  \nonumber  \\
&+&\mid\!\nabla\Psi_2\!\mid^2+2 g {\rm Re}(\nabla\Phi_2\nabla\Psi_2^*)\bigg],
\end{eqnarray}
and the perturbative Hamiltonian has the following form\cite{fixed_phase}:
\begin{eqnarray}\label{LGU}
U&=&\int d^3x\; \bigg[u_6\mid\!\Psi_6\!\mid^4+u_2\mid\!\Phi_2\!\mid^4
+\tilde u_2\mid\!\Psi_2\!\mid^4  \nonumber  \\
&+&w_1\mid\!\Psi_6\!\mid^2\mid\!\Phi_2\!\mid^2
+w_2\mid\!\Psi_6\!\mid^2\mid\!\Psi_2\!\mid^2
+w_3\mid\!\Phi_2\!\mid^2\mid\!\Psi_2\!\mid^2  \nonumber  \\
&+&h_1 {\rm Re}(\Psi_6^*\Phi_2^3)+h_2 {\rm Re}(\Psi_6^*\Psi_2^3)
+h_3 {\rm Re}(\Phi_2^2 \Psi_2^{*2}) \nonumber  \\
&+&v_1 {\rm Re}(\Psi_6^*\Phi_2\Psi_2^2)
+v_2 {\rm Re}(\Psi_6^*\Phi_2^2\Psi_2)
+v_3 \mid\!\Psi_6\!\mid^2 {\rm Re}(\Phi_2\Psi_2^*) \nonumber \\
&+&v_4 \mid\!\Phi_2\!\mid^2 {\rm Re}(\Phi_2\Psi_2^*)+
v_5 \mid\!\Psi_2\!\mid^2 {\rm Re}(\Phi_2\Psi_2^*)\bigg]
\end{eqnarray}

For the case that $r_2=\tilde r_2$, one can simply 
diagonalize the Gaussian part of the Hamiltonian 
using the transformation $\Phi_2=(\tilde\Phi_2+\tilde\Psi_2)/\sqrt 2$ 
and $\Psi_2=(\tilde\Phi_2-\tilde\Psi_2)/\sqrt 2$, and then do the
RG calculations. The RG calculations for the case 
that $\Psi_6=0$ were done by Yosefin and Domany \cite{YD85}  
in the study of the phase transitions in fully frustrated XY models

In the mean field approach, there are now four distinct 
phases for $r\neq 0$ ($\Psi_6=\Phi_2=\Psi_2=0; 
\Psi_6=0, \Phi_2\neq 0, \Psi_2\neq 0 ;
\Psi_6\neq 0, \Psi_2=\Phi_2=0;\Psi_6\neq 0, \Psi_2\neq 0,
\Phi_2\neq 0$).  
One should note that for the phases where both $\Phi_2$
and $\Psi_2$ are non-zero, we have the condition
of local stability of the free energy,
$r_2 \tilde r_2< r^2$. 
In addition, the  singularity of the propagators 
at zero wavevector (${\bf q}) = 0$)
is for $r_2 \tilde r_2=r^2$, which is the critical point of the
system.

To obtain further insight into the specific situation where
both $\Phi_2$ and $\Psi_2$ go simultaneously critical (soft),
we perform an
RG calculation.
To simplify the calculations, we consider the case that
$r_2=\tilde r_2$, so that the fields
$\Phi_2$ and $\Psi_2$ are simultaneously critical (soft),
and that they are equally coupled to the $\Psi_6$ hexatic field
($u_2=\tilde u_2$, $w_1=w_2$).  We further need to require
that the full theory, after diagonalization of the gaussian part,
is self-consistent with no new RG-generated terms. This imposes
that
$h_1=h_2$ and $v_1=v_2$.
Using the above mentioned transformation
for $\Phi_2$ and $\Psi_2$ and rescaling the fields, one can 
rewrite the Hamiltonian as,

\begin{eqnarray}\label{LG2}
&&\beta F=\int d^3x\; \bigg[\frac{r_6}{2}\mid\!\Psi_6\!\mid^2
+\frac{r_2^\prime}{2}\mid\!\tilde \Phi_2\!\mid^2
+\frac{\tilde r_2^\prime}{2}\mid\!\tilde \Psi_2\!\mid^2  \nonumber \\
&&+\frac{1}{2}\mid\!\nabla\Phi_6\!\mid^2
+\frac{1}{2}\mid\!\nabla\tilde\Phi_2\!\mid^2+
\frac{1}{2}\mid\!\nabla\tilde\Psi_2\!\mid^2
+u_6\mid\!\Psi_6\!\mid^4  \nonumber \\
&&+u_2^\prime\mid\!\tilde\Phi_2\!\mid^4 
+\tilde u_2^\prime\mid\!\tilde\Psi_2\!\mid^4
+w_1^\prime\mid\!\Psi_6\!\mid^2\mid\!\tilde\Phi_2\!\mid^2   \nonumber  \\
&&+w^\prime_2\mid\!\Psi_6\!\mid^2\mid\!\tilde\Psi_2\!\mid^2  
+w_3^\prime\mid\!\tilde\Phi_2\!\mid^2\mid\!\tilde\Psi_2\!\mid^2  
\nonumber  \\
&&+h_1^\prime {\rm Re}(\Psi_6^*\tilde\Phi_2^3)
+h_3^\prime {\rm Re}(\tilde\Phi_2^2 \tilde\Psi_2^{*2})
+v_1^\prime {\rm Re}(\Psi_6^*\tilde\Phi_2\tilde\Psi_2^2)\bigg],
\end{eqnarray}
where the new (primed) coefficients can be written in terms of
old coefficients.
To first order of $\epsilon$, the RG 
equations are given by,

\begin{eqnarray}\label{RGE2}
\frac{dr_6}{dl}&=&2 r_6+\frac{16 K_4 u_6}
{1+r_6}+\frac{4 K_4 w_1^\prime}{1+r_2^\prime}
+\frac{4 K_4 w_2^\prime}{1+\tilde r_2^\prime}  \nonumber  \\
\frac{dr_2^\prime}{dl}&=&2 r_2^\prime+\frac{16 K_4 u_2^\prime}{1+r_2^\prime}
+\frac{4 K_4 w_1^\prime}{1+r_6}
+\frac{4 K_4 w_3^\prime}{1+\tilde r_2^\prime}  \nonumber  \\
\frac{d\tilde r_2^\prime}{dl}&=&2 \tilde r_2^\prime
+\frac{16 K_4 \tilde u_2^\prime}{1+\tilde r_2^\prime}
+\frac{4 K_4 w_2^\prime}{1+r_6}
+\frac{4 K_4 w_3^\prime}{1+r_2^\prime}  \nonumber  \\
\frac{du_6}{dl}&=&\epsilon u_6-40 K_4 u_6^2
-2 K_4 w_1^{\prime 2}-2 K_4 w_2^{\prime 2} \nonumber \\ 
\frac{du_2^\prime}{dl}&=&\epsilon u_2^\prime-40 K_4 u_2^{\prime 2}
-2 K_4 w_1^{\prime 2}-2 K_4 w_3^{\prime 2}
-9 K_4 h_1^{\prime 2} \nonumber \\
&-&2 K_4 h_3^{\prime 2}   \nonumber \\
\frac{d\tilde u_2^\prime}{dl}&=&\epsilon\tilde u_2^\prime
-40 K_4 \tilde u_2^{\prime 2}-2 K_4 w_2^{\prime 2}
-2 K_4 w_3^{\prime 2}-2 K_4 h_3^{\prime 2}  \nonumber \\
&-&K_4 v_1^{\prime 2}  \nonumber \\
\frac{dw_1^\prime}{dl}&=&\epsilon w_1^\prime
-16 K_4 u_6 w_1^\prime-16 K_4 u_2^\prime w_1^\prime
-8 K_4 w_1^{\prime 2}  \nonumber \\
&-&4 K_4 w_2^\prime w_3^\prime
-18 K_4 h_1^{\prime 2}-2 K_4 v_1^{\prime 2} \nonumber \\
\frac{dw_2^\prime}{dl}&=&\epsilon w_2^\prime
-16 K_4 u_6 w_2^\prime-16 K_4 \tilde u_2^\prime w_2^\prime
-8 K_4 w_2^{\prime 2} \nonumber \\
&-&4 K_4 w_1^\prime w_3^\prime-4 K_4 v_1^{\prime 2}  \nonumber \\
\frac{dw_3^\prime}{dl}&=&\epsilon w_3^\prime-16 K_4 u_2^\prime w_3^\prime
-16 K_4 \tilde u_2^\prime w_3^\prime-8 K_4 w_3^{\prime 2}  \nonumber \\
&-&4 K_4 w_1^\prime w_2^\prime-16 K_4 (h_3^\prime)^ 2
-4 K_4 v_1^{\prime 2}  \nonumber \\
\frac{dh_1^\prime}{dl}&=&\epsilon h_1^\prime-24 K_4 h_1^\prime u_2^\prime
-12 K_4 w_1^\prime h_1^\prime-4 K_4 v_1^\prime h_3^\prime   \nonumber \\
\frac{dh_3^\prime}{dl}&=&\epsilon h_3^\prime-8 K_4 u_2^\prime h_3^\prime
-8 K_4 \tilde u_2^\prime h_3^\prime-16 K_4 w_3^\prime h_3^\prime  \nonumber \\
&-&6 K_4 h_1^\prime v_1^\prime   \nonumber \\
\frac{dv_1^\prime}{dl}&=&\epsilon v_1^\prime-8 K_4 v_1^\prime \tilde u_2^\prime
-4 K_4 w_1^\prime v_1^\prime-8 K_4 w_2^\prime v_1^\prime   \nonumber \\
&-&8 K_4 w_3^\prime v_1^\prime-12 K_4 h_1^\prime h_3^\prime .
\end{eqnarray}


We have found a number of FPs for the above RG equations that
fulfill the condition 
($r_2^*=\tilde r_2^*$, and for which
$u_2^*=\tilde u_2^*$, $w_1^*=w_2^*$, $h_1^*=h_2^*$, $v_1^*=v_2^*$
and $v_4^*=v_5^*$).
Some of these are given in the Table~I, where each column correspond to a
different FP (the nontrivial mixed herringbone-hexatic
FP found in Section IIa occurs here as well, but is again
unstable).  As in the simplest case of Hamiltonian (2.1),
none of the FP correspond to a nontrivial stable fixed point: each
fixed point (column) shows more than two positive eigenvalues. 
Given the complexity of those nonlinear equations we cannot be
100\% sure that we have found all the (unstable) FP in the
$(u_6,u_6^\prime,\tilde u_2^\prime, w_1^\prime, w_3^\prime,
h_1^\prime, h_3^\prime, v_1^\prime)$
plane. However, a numerical investigation of the RG flow in
that plane starting from a large number of initial values for
the coupling parameters always failed to converge towards a stable 
(attractive) fixed point.
Therefore, we
reach the same conclusion obtained in the previous section, namely that
of a fluctuation-driven first order transition for this expanded symmetry
Hamiltonian, and the stabilization of novel fixed point in $N$ coupled
two vector models for $N>2$ does not appear to occur in this generalized
Hamiltonian
due to the
extra coupling parameters.

\newpage

\begin{center}

{\bf Table I: The FPs of RG equations for generalized Hamiltonian.}

\vspace{0.5cm}

\begin{tabular}{ccccccccc}
\hline
$r_6^{\prime*}/\epsilon$             & -1/5 & -3/11 & -2/7 & 0 & -0.2761944898 & -0.2583527792 & -0.2727272730 \\
$r_2^{\prime*}/\epsilon$             & -1/5 & -3/11 & -2/7 & -1/5 & -0.2761944898 & -0.2756996328 & -0.2727272730 \\
$\tilde r_2^{\prime*}/\epsilon$      & -1/5 & -3/11 & -2/7 & -1/5 & -0.2566490108 & -0.2756996328 & -0.2727272724 \\
$u_6^{\prime*}(K_4/\epsilon)$        & 1/40 & 1/44 & 1/56 & 0 & 0.0202792491 & 0.0237809941 & 0.0227272727 \\
$u_2^{\prime*}(K_4/\epsilon)$        & 1/40 & 1/44 & 1/56 & 0.0125 & 0.0202792491 & 0.0194412033 & 0.0170454546 \\
$\tilde u_2^{\prime*}(K_4/\epsilon)$ & 1/40 & 1/44 & 1/56 & 0.0125 & 0.0238702512 & 0.0194412033 & 0.0170454546 \\
$w_1^{\prime*}(K_4/\epsilon)$        & 0 & 1/44 & 1/28 & 0 & 0.0405584982 & 0.0170262066 & 0.0227272727 \\
$w_2^{\prime*}(K_4/\epsilon)$        & 0 & 1/44 & 1/28 & 0 & 0.0164217503 & 0.0170262066 & 0.0227272727 \\
$w_3^{\prime*}(K_4/\epsilon)$        & 0 & 1/44 & 1/28 & 0.05 & 0.0164217503 & 0.0430587966 & 0.0454545454 \\
$h_1^{\prime*}(K_4/\epsilon)$        & 0 & 1/44 & 0 & 0 & 0 & 0 & 0 \\
$h_3^{\prime*}(K_4/\epsilon)$        & 0 & 1/44 & 0 & $\pm$ 0.025 & 0 & $\pm$ 0.0041763898 & $\pm$ 0.0113636364 \\
$v_1^{\prime*}(K_4/\epsilon)$        & 0 & 1/44 & 0 & 0 & 0 & 0 & 0 \\
\hline
\end{tabular}

\end{center}

\vspace{1cm}

\section{Discussion}

In Section II. A, we found that there are two ``new'' FPs
missed in Ref. \cite{BA82}. However, those FPs are unstable, 
and we reach the same final conclusion as Bruinsma and Aeppli, namely
that the
SmA$-$HexB transition is driven first order by fluctuations in the
BA Hamiltonian.
We also 
discussed a slightly modified simple model that considers not only
the hexatic and herringbone order, but also one that
involves the local two-fold deformation of the
bond correlations induced by the herringbone correlations.
We assumed that this deformation can 
be represented
by a two-fold symmetry order parameter, as
in the case of herringbone order, and wrote
the free energy based on symmetry arguments.
We were not able to find a new stable FP
which could possibly result in unconventional (``new'')
second order critical exponents.

It is not clear to us
in what directions to pursue the paradoxical puzzle
of ``new universality'' (non $3D$ XY transition) in hexatic liquid crystal
materials. 
In the work presented here, 
we have found a novel ``mixed'' hexatic-herringbone
fixed point in the theory, but which is unstable to first
order in $\epsilon$. One note that the two positives
eigenvalues $y_5$ and $y_6$ in Eq. 2.4 are only very slightly
positive for $\epsilon =1$.
This observation may open the 
possibility that in a calculation that includes
hexatic-herringbone
coupling, there is no stable fixed point to lowest order in
$\epsilon$, but that the fixed point may be stabilized in
a theory that goes beyond an
$O(\epsilon^1)$ calculation.
This is what happens, for example,
in the normal to superconducting
phase transition where the $\epsilon$ expansion to
order $\epsilon^1$ predicts a fluctuation-driven
first order phase transition\cite{HLM} while theoretical
arguments and Monte Carlo simulations strongly argue for
a  second order (inverted 3D XY) phase transition\cite{Dasgupta}.

\newpage

.
\vspace{8.5cm}

Having said that, one
should note that there are
experiments on liquid
crystal materials that do not display any herringbone correlations
~\cite{Jin,Gorecka} but still show a SmA$-$HexB transition with
exponents that differ from the critical behavior expected to a
$2D$~\cite{Jin} or $3D$\cite{Gorecka} XY critical behavior.
That may suggest that the whole idea of hexatic$-$herringbone
interactions is a red herring.
Another possibility is that
of (more subtle) `hidden'
order parameter(s) distinct from the herringbone order characterizes the 
SmA$-$HexB transition in real materials, and that the
coupling between
this hidden order parameter and the hexatic order 
parameter $\Psi_6$ produces a novel stable fixed point.
Clearly, more experimental studies are 
needed to shed light on this problem.
In particular, 
high resolution scattering experiments
would seem necessary to search for extended short-range 
correlation in either molecular correlations
or distortion of hexagonal co-ordination to shed
some light on what such `hidden' order parameter(s)
may be.

We finally note that a related (unconventional 
critical behavior) situation arises in the 
context of layered systems of smectic liquid
crystals studied by Defontaines and Prost
\cite{DP93}. These authors
 have argued that critical points
that do not involve any symmetry change can define
a set of new universality classes in layered systems.
 Possibly,
considerations of some features of the Defontaines and
Prost theory may be useful
in further investigations of the SmA$-$HexB
problem.

We hope that our work and reinvestigation of the long
standing SmA$-$Hex-B transition in smectic liquid crystal
materials will motivate further theoretical, numerical and
experimental investigations of this very interesting problem.

\section{Acknowledgements}

We thank Ian Affleck, Igor Herbut and
and Mehran Kardar for useful 
discussions. M. G. thanks C. C. Huang for stimulating his 
interest in the problem of the critical behavior
of the SmA$-$HexB transition.
This research has been funded by NSERC of Canada.  
M. G. acknowledges the Research Corporation and the 
Province of Ontario for generous financial support.


\begin{thebibliography}{99}
\bibitem{cryE_crystal}
The parametrization of the
crystalline E phase should also include a 
third order parameter: the amplitude of the Fourier
transform of the mass density at the Bragg reciprocical
wave vector.
\bibitem{fixed_phase} It is worth noting that we implicitely make the assumption here that the 
effective symmetry-breaking fields that couple the $\Psi_6$, $\Psi_2$ and $\Phi_2$ 
fields are assumed to have an homogeneous and constant phase relationship arbitrarily set to zero
in spirit to the theories used in Refs.~\cite{BA82,locked-hexatic_nematic}.
In other words, the Ginzburg-Landau-Wilson free energies for the smetic-A$-$hexatic-B problem
are not local gauge theories.
\end{thebibliography}
\end{document}